\input phyzzx.tex

\twelvepoint


\def\m{\hat m}
\def\n{\hat n}
\def\p{\hat p}
\def\q{\hat q}

\def\s{\hat s}
\def\bs{\bar s}
\def\bu{\bar u}
\def\a{\hat a}
\def\b{\hat b}
\def\c{\hat c}
\def\d{\hat d}
\def\e{\hat e}


\REF\SW{N. Seiberg and E. Witten, Nucl. Phys. {\bf B426} (1994) 19, 
hep-th/9407087; Nucl. Phys. {\bf B431} (1994) 484, hep-th/9408099}
\REF\HStelleW{P. Howe, K. Stelle and P. West, Phys. Lett. {\bf 124B} (1983) 55.
For a review of perturbative results in rigid supersymmetric theories see
P. West,  {\it Supersymmety and Finiteness}, in 
Proceedings of the
1983 Shelter Island II Conference on Quntum Field Theory and
Fundamental Problems of Physics, edited by R. Jackiw, N. Kuri, S.
Weinberg and E. Witten (M.I.T. Press)}
\REF\Seiberg{N. Seiberg, Phys. Lett. {\bf 318B} (1993) 469}
\REF\MO{C. Montonen and D. Olive, Phys. Lett. {\bf 72B} (1977) 117}
\REF\Confirm{N. Dorey, V. V. Khoze and M. P. Mattis, Phys. lett. {\bf B388}
(1996) 324, hep-th/9607066; Phys. Rev.  {\bf D54} (1996) 2921; Phys. Rev. {\bf 
D54} (1996) 
7832, hep-th/9607202; 
K. Ito and N. Sasakura, Phys. Lett. {\bf B382} (1996) 95, hep-th/9602073; 
Nucl. Phys. {\bf B484} (1997) 141, hep-th/9608054; 
A. Yung, Nucl. Phys. {\bf B485} (1997) 38, hep-th/9605096;
H. Aoyama, T. Harano, M. Sato and S. Wada, Phys. lett. {\bf B388} (1996) 331, 
hep-th/9607076;
T. Harano and M. Sato, Nucl.Phys. {\bf B484} (1997) 167, hep-th/9608060}
\REF\Matone{M. Mantone, Phys. Lett. {\bf 357B} (1995) 342, hep-th/9506102}
\REF\HW{P. Howe and P. West,  Nucl. Phys. {\bf B486} (1997) 425, 
hep-th/9607239}
\REF\MOS{M. Magro, L. O'Raifeartaigh and I. Sachs,
{\it Seiberg-Witten Effective Lagrangian from
Superconformal Ward Identities}, hep-th/9704027}
\REF\AF{P. Argyres and A. Faraggi, Phys. Rev. Lett. {\bf 74} (1995) 3931,
hep-th/9411057; S. Klemm, W. Lerche, S. Theisen and S. Yankielowicz,
Phys. Lett. {\bf 344B} (1995) 169, hep-th/9411048}
\Ref\KV{S. Kachru and C. Vafa, Nucl. Phys. {\bf B450} (1995) 69, 
hep-th/9505105; S. Kachru, A. Klemm, W. Lerche, P. Mayr and C. Vafa,
Nucl. Phys. {\bf B359} (1995) 537, hep-th/9508155;
A. Klemm, W. Lerche, P. Mayr, C. Vafa and N. Warner, Nucl. Phys. 
{\bf B477} (1996) 746,
hep-th/9604034}
\REF\Witten{E. Witten, {\it Solutions of Four Dimensional Field Theories
via M Theory}, hep-th/97003166}
\REF\HSW{P.S. Howe, E. Sezgin and P.C. West, Phys. Lett. {\bf B399}
(1997) 49, hep-th/9702008}
\REF\others{M. Perry and J.H. Schwarz, Nucl. Phys. {\bf B489} (1997) 47,
hep-th/9611065; M. Aganagic, J. Park, C. Popescu, and J. H. Schwarz,
{\it Worldvolume action of the M-theory fivebrane},
hep-th/9701166;
I. Bandos, K Lechner, A. Nurmagambetov, P. Pasti and D. Sorokin, and M.
Tonin, {\it Covariant action for the super fivebrane of M-theory},
hep-th/9701149}
\REF\one{P.S. Howe, N.D. Lambert and P.C. West, {\it The Self-Dual String
Soliton}, hep-th/9709014}
\REF\three{P.S. Howe, N.D. Lambert and P.C. West, {\it The Threebrane
Soliton of the M-Fivebrane}, hep-th/9710033}

\pubnum={KCL-TH-97-55\cr hep-th/9710034}
\date{September 1997}

\titlepage

\title{\bf Classical M-Fivebrane Dynamics and Quantum $N=2$ Yang-Mills}

\centerline{P.S. Howe}
\centerline{N.D. Lambert}

\centerline{and}

\centerline{P.C. West\foot{phowe, lambert, pwest@mth.kcl.ac.uk}}
\address{Department of Mathematics\break
         King's College, London\break
         England\break
         WC2R 2LS\break
         }

\abstract
We obtain the complete quantum Seiberg-Witten effective action for 
$N=2$ supersymmetric $SU(N)$ Yang-Mills
theory from the classical M-fivebrane equations of motion with  $N$ 
threebranes moving in its worldvolume.


\chapter{Introduction}

The  remarkable solution of  Seiberg and Witten [\SW] for the
complete non-perturbative chiral part of the effective action  of
$SU(2)$
$N=2$ Yang-Mills theory spontaneously broken to $U(1)$ represents one
of the few examples in quantum field theory where the complete quantum
effects are known.  The  result was derived  using essentially two
inputs; the first was the complete  perturbative  result for this
action which had been known  for many years [\HStelleW,\Seiberg] and 
the second was
an application of electromagnetic duality [\MO]. Since only the first
of these results is well established there have been a number of
attempts to give alternative derivations of the Seiberg-Witten
effective action. Some of the terms in the
effective action have been confirmed using instanton corrections [\Confirm].
It  has also been observed that the Seiberg-Witten effective action
obeys a particular non-trivial relation [\Matone], which has been 
shown to be a consequence of  the anomalous  $N=2$
superconformal Ward identities of a  spontaneously broken $N=2$
Yang-Mills theory [\HW]. 
It has since been
argued [\MOS] that this relation  implies the full effective
action.

One of the most mysterious features of the Seiberg-Witten
effective action is the way that it is related to an
associated Riemann surface. In particular it was shown how the effective  
action could be constructed given the Riemann surface using a specific recipe.
However, how this surface arises naturally in the theory was
not apparent.  Shortly after the
appearance of references [\SW]  Riemann surfaces that
were thought to correspond to the spontaneously broken
gauge groups $SU(N)$ were proposed [\AF].

An alternative approach has been to attempt to derive the Seiberg-Witten
results from a string theory. 
It was found in [\KV] that the Seiberg-Witten curves occured in
Calabi-Yau Compactifications of type IIB string theory and furthermore, 
from the M theory perspective, these could be interpreted as a single
fivebrane wrapped on a Riemann surface.  More recently,
Witten [\Witten] considered configurations of intersecting NS-fivebranes and 
D-fourbranes in  IIA string theory. He argued that the spontaneously broken
$N=2$ Yang-Mills theory appeared on the parallel D-fourbranes. By
embedding this picture in M theory  he argued that the configuration
could be represented by a single M-fivebrane and  was able to show that
the Riemann surfaces which  arise in not only  the
$SU(2)$ theory, but also in its
$SU(N)$ generalisations, appear in a very natural way. However,
in this work the connection between the M-fivebrane dynamics and
the final Seiberg-Witten effective action remained obscure.

In this paper, we shall use the  M-theory fivebrane
classical dynamics to derive the Seiberg-Witten effective action.
In particular we  use the M-fivebrane dynamics as formulated in
[\HSW], although there are other formulations [\others].
It was recently
shown that the M-fivebrane admits onebrane [\one] and threebrane solutions
[\three] within it.  We will consider  a single M-fivebrane on which some
threebranes are moving. The zero modes of this configuration
correspond to the modes of the spontaneously broken $N=2$ Yang-Mills
theory which can be viewed as living on the four-dimensional worldvolume  
of a  threebrane. It is a
consequence of the Bogomoln'yi  condition for the threebranes that the
M-fivebrane can be viewed as being wrapped on a Riemann surface which is
itself embedded in a four-dimensional space. This latter space is
composed of the two dimensions of the M-fivebrane transverse to the
threebranes and the active two of the five transverse dimensions of
the fivebrane embedded in M theory. We consider the classical
fivebrane action for threebrane configurations in which the
zero modes are allowed to depend on the threebrane coordinates. We 
show that it is
precisely the same as the full non-perturbative Seiberg-Witten
effective action for spontaneously broken $SU(N)$ Yang-Mills gauge
theory.  In practice  we only carry out this calculation
for the scalars, but  it follows from $N=2$ supersymmetry
that it holds
for the full action involving the fermions and vectors.


\chapter{The Threebrane Effective Action}

The M theory fivebrane  has a six-dimensional $(2,0)$ tensor multiplet of 
massless fields on its worldvolume.
The classical equations of motion  in the absence
of fermions and background fields are [\HSW]
$$\eqalign{
G^{\m\n}\nabla_{\m}\nabla_{\n} X^{a'} &= 0 \ , \cr
G^{\m\n}\nabla_{\m}H_{\n\p\q} & = 0 \ ,\cr}
\eqn\eqom
$$
where the worldvolume indices are $\m,\n=0,1,...,5$ and the transverse indices
are $a',b'=6,7,8,9,10$. In \eqom\ $\nabla$ is the Levi-Civita connection of
the metric $g_{\m\n}=\eta_{\m\n} + \partial_{\m}X^{a'}\partial_{\n} X^{a'}$.
The three form $H_{\m\n\p}$ is closed but not self-dual. Rather, in 
the vielbein frame defined by $E_{\a}^{\ \m}$, $\a,\b=0,1,...,5$, it is
related to the self-dual three form $h_{\a\b\c}$ via $H_{\a\b\c} 
= m_{\a}^{\ \d}m_{\b}^{\ \e}h_{\d\e\c}$, where 
$m_{\a}^{\ \b} = \delta_{\a}^{\ \b} -2h_{\a\c\d}h^{\b\c\d}$. 
The vielbein $E_{\a}^{\ \m}$ is associated to the metric $G^{\m\n}$
through $G^{\m\n} = E_{\a}^{\ \m}E_{\b}^{\ \n}\eta^{\a\b}$ but  the metric
$g^{\m\n}$ is  obtained from the vielbein 
$e_{\a}^{\ \m} =(m^{-1})_{\a}^{\ \b}E_{\b}^{\ \m}$.
We refer the reader to references [\HSW,\one] 
for more details of the formalism and notation. 

Using the solution of [\three] and adopting the notation of
[\Witten] we  consider a multi-threebrane solution
which lies in the $(x^0,x^1,x^2,x^3)$ plane. We take only two transverse
fields $X^{6}$ and $X^{10}$ to be active and assume that $X^{10}$ is a compact 
dimension of radius $R$. The  other three scalars are  
constant and the three form vanishes. First let us
consider the dependence of the fields on the M-fivebrane coordinates 
transverse 
to the threebranes, $x^4$ and $x^5$. This solution preserves 
half of the six-dimensional supersymmetries if [\three]
$$
\partial_4 X^{6} = \partial_5 X^{10} \ ,\ \ \ \ \ 
\partial_5 X^{6} = -\partial_4 X^{10}\ . 
\eqn\halfsusy
$$
This leads to an 
$N=2$ vector mulitplet on the four-dimensional worldvolume of the
threebranes [\three].
Adopting the complex notation $s = (X^{6} + iX^{10})/R$ and 
$z = \Lambda^2(x^4+ ix^5)$, where $\Lambda$ is a mass scale,
we recognise \halfsusy\ as the 
Cauchy-Riemann equation. Thus  $s$ is a complex function of  $z$ only. 
Furthermore any choice of this function will 
solve the field equations [\three].
The threebranes can be then thought of as an M-fivebrane in an 
eight-dimensional
space with coordinates $x^0,...,x^5,X^6,X^{10}$, in which $x^0,...,x^3$ are
flat ${\bf R}^4$ and $x^4,x^5,X^6,X^{10}$ form a non-trivial four-dimensional
space $Q$. The M-fivebrane field equations in the presence of these 
threebranes  then imply that the M-fivebrane is 
wrapped on a Riemann surface $\Sigma$, which is embedded  in the 
four-dimensional space $Q$. The volume of this Riemann surface is set by the 
scale $R^2$. Given
the geometrical construction of the Riemann surface presented  here, it is
perhaps more natural to assign $z$ the dimensions of length. However we have
assigned  $z$ the dimensions of mass in order to make contact with the 
literature on the Seiberg-Witten solution.

Following [\Witten] we define $t=e^{-s}$ and consider a
threebrane configuration defined by
$$
F(t,z) = 0 \ ,
\eqn\Fdef
$$
where $F$ is a complex polynomial. In order to make contact with [\Witten]
let us consider the  
IIA picture in ten dimensions obtained by taking the small $R$ limit. In this
case there are D-fourbranes in the $(x^0,x^1,x^2,x^3,X^6)$ plane, located 
at the 
roots of $F(\ {\cdot},z)=0$ 
and also NS-fivebranes in the $(x^0,x^1,x^2,x^3,x^4,x^5)$ plane,
located at the roots  of $F(t,{\cdot}\ )=0$ [\Witten]. The threebrane is
the intersection of the D-fourbranes with the NS-fivebranes.
The restriction
to a polynomial $F$ then   ensures that there are only a finite number of 
branes. 

A simple 
example of such a configuration is  a collection of $N$ threebranes  at
the positions $d_i$, $i=1,...,N$, with charges $q_i$ [\three]
$$
s = s_0-\sum_i q_i\ln (z-d_i)\ ,
\eqn\sexample
$$
where $s_0$ is an aribitrary constant which we set to zero.
The corresponding surface is
$$
F(t,z)\equiv t^2 - \prod_i (z-d_i)^{2q_i} = 0 \ ,
\eqn\Fexample
$$ 
where $t$ has been
suitably rescaled by $\Lambda$.
For integer  values of $q_i$ this is a singular 
Riemann surface, with degenerate
roots, while if $q_i = {1\over2}$ it is non-singular. In fact we will use
a more general threebrane configuration below.

The scalar fields in the resulting four-dimensional theory in ${\bf R}^4$
are the positions of the threebranes $d_i$. We therefore allow $s$ to
be function of $x^{\mu},\  \mu=0,1,2,3$ by letting the locations of the
threebranes become $x^{\mu}$ dependent.  As seen from the M-fivebrane this
corresponds to letting the moduli of the Riemann Surface depend on $x^{\mu}$. 
We now wish to evaluate
the dynamics for this configuration. For simplicity we will just consider
the scalar fields with the three form field of the M-fivebrane vanishing,
or equivalently with the  vector fields in the four-dimensional effective
theory set to zero. 

For such a configuration the equation of motion is
$g^{\m\n}\nabla_{\m}\nabla_{\n}s = 0$
and can be derived from the usual brane action
$$
I_5 = \int d^6 x \sqrt{-\det g_{\m\n}} \ ,
\eqn\braneaction
$$
where, for a threebrane  configuration,  
$$
g_{\m\n} = \eta_{\m\n} 
+ {1\over2}(\partial_{\m}s\partial_{\n}\bs+\partial_{\m}\bs\partial_{\n}s)\ .
\eqn\gsimple
$$
One can then evaluate the determinant in \braneaction\ to find
$$
-\det g_{\m\n} = (1 + {1\over 2}\partial_{\m}s\partial^{\m}\bs)^2 
- {1\over4}|\partial_{\m}s\partial^{\m}s|^2\ .
\eqn\detis 
$$
In the above expressions and below we have suppressed factors of $R$ which
appear with $s$.
To obtain the low energy effective action for the threebrane we will only
keep terms of second order in the derivatives $\partial_{\mu}$. Using the
equation of motion of of $s$ one readily finds 
$$
I_5 = {1\over 2}\int d^6 x  \partial_{\mu}s\partial^{\mu}\bs \ .
\eqn\Ione
$$
Let us now denote the moduli of the Riemann surface by $u_i = u_i(x)$. The 
above action becomes
$$
I_5 = {1\over 4 i} \int d^4x\partial_{\mu}u_i\partial^{\mu}\bu_j 
\int_{\Sigma}\omega^i\wedge {\bar \omega}^j\ ,
\eqn\Itwo
$$
where $\omega^i = -\partial s /\partial u_i\ dz$. To evaluate this action
further we must adopt a particular form for the threebrane configuration and 
hence the Riemann surface $\Sigma$.

We wish to consider 
threebrane 
configurations
which are everywhere smooth, so that the associated Riemann surface is
also smooth,  but which asymptotically take the form
\sexample . To this end for the rest of this paper we shall restrict our  
attention to functions $F$ of the form
$$
F(t,z) = t^2 - 2B(z)t + \Lambda^{2N} = 0 \ ,
\eqn\Ftwo
$$
where $B(z)$ is a polynomial of degree $N\ge 2$ and $t$ has again been
suitably rescaled by $\Lambda$. 
However, it is also possible to consider more complicated threebrane
configurations. The choice \Ftwo\ leads to the scalar function
$$
s = s_0-\ln\left(B +\sqrt{Q} \right)\ ,
\eqn\s
$$
where we  have introduced the polynomial $Q = B^2(z) - \Lambda^{2N}$ and 
$s_0$
is a constant determined by $N$ and $\Lambda$.
A sufficiently general parameterisation of $B(z)$ in terms of the $u_i$'s is 
$$
B(z) = z^{N} + u_{N-1} z^{N-2} + u_{N-2} z^{N-3}+...+u_{1}\ .
\eqn\Bdef
$$
Note that only the $z^{N-1}$ term is missing from \Bdef. This is because, as 
noted in [\Witten], such a term would cause the integral over the Riemann
surface in  \Ione\ to diverge. 

Lets us pause for a moment to consider the physical interpretation of this 
choice
of $F$ in terms of the threebranes. If $\Lambda=0$ then this solution is
described by \sexample\ with all the $q_i=1$ and the positions $d_i$ of the
threebranes are the $N$ roots of $B$. From this point of view a 
$z^{N-1}$ term in \Bdef\ corresponds to the centre of mass coordinate for
the threebranes (its coefficient $u_{N}$ is the sum of the roots). 
Low energy motion in this moduli would produce  an infinite contribution
to the action \Ione\ and so has been frozen out. For a non-zero value of  
$\Lambda$ the Riemann
surface is generically non-singular, but the picture in
terms of the simple threebranes of \sexample\ is slightly obscured. 
For $|z|>>\Lambda$ we
again see $N$ distinct threebranes located at $d_i$ 
but for finite $z$ this is not the case. In fact one can see from \Ftwo\ that
there are no points $t=0$, corresponding to the singular core of a threebrane.
In terms of the ten-dimensional type IIA picture  \Ftwo\  corresponds to
two NS-fivebranes with $N$ D-fourbranes suspended between them [\Witten].

We now  evaluate the effective action \Ione . As explained in 
[\Witten], the Riemann surface can be defined by
$$
{\tilde t}^2 = Q(z) = B^2(z) - \Lambda^{2N} \ ,
\eqn\ttilde
$$
where ${\tilde t} = t - B$. It is a straightforward calculation to see that
$$
\omega^i = -{\partial s\over \partial u_i} dz 
= {z^{i-1}\over \sqrt{Q}} = \lambda^i\ ,
\eqn\holo
$$
where $\lambda^i$ is the $i$th holomorphic form of the Riemann surface. 
Using
the Riemann bi-linear relation we may express the effective action as
$$
I_5 = {1\over 4i} \int d^4 x \ \partial_{\mu}u_i\partial^{\mu}\bu_j
\sum_{k=1}^{N-1}\left(\int_{A_k}\lambda^i \int_{B^k}{\bar \lambda}^j
- \int_{A_k}{\bar \lambda}^j\int_{B^k}{\lambda}^i \right) \ ,
\eqn\Ithree
$$
where $A_k$ and $B^k$ are a basis for 
the  $a$ and $b$ cycles on the Riemann surface.

Next we note that 
$\partial_{\mu}u_i\lambda^i = \partial_{\mu}\lambda_{SW}$,
where we have introduced the Seiberg-Witten differential $\lambda_{SW}$ 
which satisfies [\SW]
$$
{\partial \lambda_{SW}\over \partial u_i} = \lambda^i  \ .
\eqn\swdef
$$
Therefore if we define the new scalar modes
$$
a_i = \int_{A_i} \lambda_{SW} \ , \ \ \ \ \ \ 
a_D^i = \int_{B^i} \lambda_{SW} \ ,
\eqn\aaddef
$$ 
we obtain the  effective action
$$\eqalign{
I_{5} &= {1 \over 4i}\int d^4 x\ \left( 
\partial_{\mu}a_k\partial^{\mu}{\bar a}_D^k 
-\partial_{\mu}{\bar a}_k\partial^{\mu}a_D^k \right)\ ,\cr
&=-{1\over 2}{\rm Im}\left(\int d^4x\ \partial_{\mu}{\bar a}_i
\partial^{\mu}{a}_j
\tau^{ij}\right)
\ .\cr}
\eqn\SWaction
$$
In \SWaction\ we have introduced $\tau^{ij} = \partial a_D^i /\partial a_j$, 
which on the general grounds of $N=2$ supersymmetry may be obtained from 
the holomorphic prepotential ${\cal F}$ as 
$\tau^{ij} = \partial^2 {\cal F}/\partial a_i\partial a_j$ [\SW]. 
It was shown in [\Witten] that the brane configuration we are
considering produces the correct Riemann surface and hence 
the correct Seiberg-Witten differential for $N=2$ $SU(N)$ Yang-Mills [\AF].
Thus we have
arrived at the scalar part of the full Seiberg-Witten effective action
for $N=2$ $SU(N)$ Yang-Mills theory. 
Furthermore as a consequence of $N=2$ supersymmetry  ${\cal F}$ uniquely 
determines the 
entire low energy effective theory,
including the terms that we would have obtained by considering the 
fermionic and  and vector zero modes.  Thus the effective action for all
of the threebrane zero modes is the complete Seiberg-Witten effective action.
We leave the details of this  calculation to be performed elsewhere.

Finally we would like to point out that in obtaining the Seiberg-Witten
effective action we have discarded various higher derivative terms in
\braneaction . If we include these the full action for the scalars is
$$\eqalign{
I_5 &= \int d^4x\int_{\Sigma}d^2z \sqrt{\left(1 + 
\Lambda^{4}R^2|\partial_zs|^2 + 
{1\over 2}R^2\partial_{\mu}s\partial^{\mu}\bs\right)^2 - 
{1\over 4}R^4|\partial_{\mu}s\partial^{\mu}s|^2}
\ ,\cr
& = I_{SW} - {R^4\over 8}\int d^4x\int_{\Sigma} d^2z\ 
{|\partial_{\mu}s\partial^{\mu}s|^2\over 
1 + R^2\Lambda^4|\partial_zs|^2 }
+ {\cal O}(|\partial_{\mu}s|^6) \ ,}
\eqn\higher
$$
where $I_{SW}$ is the Seiberg-Witten effective action \SWaction\ and we have
included all the factors of $R$.
By writing $\partial_{\mu}s = \lambda^i\partial_{\mu}u_i$ and continuing
the  expansion in \higher\ one obtains an infinite series of higher
derivative terms involving powers of $\lambda^i$. 
It is natural to suppose that these terms correspond to some of the higher 
derivative corrections to the Seiberg-Witten effective action. Note
that the terms in \higher\ have a particular form; where
each $u_i$ has one and only one derivative acting on it. One may  
hope that these are all of the higher derivative terms of this type.


\chapter{Discussion}

In this paper we have considered the M theory  fivebrane with
threebranes moving within it. We evaluated the classical M-fivebrane
equations of motion  for this field configuration  when the zero
modes of the threebrane solutions are taken to depend on $x^{\mu}$, the 
four-dimensional coordinates of  threebrane worldvolume. The resulting 
four-dimensional theory corresponds to a spontaneously broken
$N=2$ gauge theory and the corresponding low energy effective action  is that
given by Seiberg and Witten. Hence we have derived the complete
{\it quantum} non-perturbative Seiberg-Witten effective action from
the  {\it classical} dynamics of the M theory fivebrane.

Given our conventional understanding of the relationship between
classical and quantum theories it is surprising to see such a detailed
and complete connection. Although M theory has only one scale $M_{Planck}$,
the solution for the threebranes introduces two more scales $R$ and 
$\Lambda$, which occur as integration constants. 
Reintroducing $\hbar$ we would
find that it occurs in combination with these integration constants. Since
these constants and so $\hbar$ do not play the r{\^o}le of a perturbative
parameter in the derivation we have given, from the M theory perspective it 
is perhaps not so surprising to
have found the full non-perturbative Seiberg-Witten effective
action. 

We note that we have only obtained the classical 
effective action for
the threebrane collective coordinates, which one must then start to quantise.
This corresponds to the fact that the Seiberg-Witten effective action 
is obtained by
integrating out the massive modes in the path integral. However in the full
quantum theory one is still left with the integral over the 
massless $U(1)^{N-1}$ modes. Along these lines one may also wonder how the 
non-Abelian structure of
$SU(N)$ Yang-Mills can be seen from the M-fivebrane perspective.

Finally, although we have used the constraints of  $N=2$
supersymmetry  to deduce the full Seiberg-Witten effective action 
from only its scalar part, this was not crucial. We could also have  explicitly
derived these
terms by considering the collective coordinates of the  three form.  Thus
one may wonder if the proceedure considered here 
can be applied to situations    with less
supersymmetry, perhaps leading to effective actions for some $N=1$ 
supersymmetric theories.

P.C.W. would like to thank D.I. Olive for discussions. 

\refout

\end